# Extreme sub-wavelength magneto-elastic electromagnetic antenna implemented with multiferroic nanomagnets[1]

J. L. Drobitch[1], A. De[2], K. Dutta[2], P. Pal[2], A. Adhikari[2], A. Barman[2] and S. Bandyopadhyay[1]

[1]Dept. of Electrical and Computer Engineering, Virginia Commonwealth University, Richmond, VA 23284, USA

[2]Dept. of Condensed Matter Physics and Material Science, S. N. Bose National Center for Basic Sciences, JD Block, Sector III, Salt Lake, Kolkata 700 106, India

## Abstract

Antennas typically have emission/radiation efficiencies bounded by $A/\lambda^2$ $\left(A<\lambda^2\right)$ where $A$ is the emitting area and $\lambda$ is the wavelength of the emitted wavelength. That makes it challenging to miniaturize antennas to extreme sub-wavelength dimensions. One way to overcome this challenge is to actuate an antenna not at the resonance of the emitted wave, but at the resonance of a different excitation that has a much shorter wavelength at the same frequency. We have actuated an electromagnetic (EM) antenna with a surface acoustic wave (SAW) whose wavelength is about five orders of magnitude smaller than the EM wavelength at the same frequency. This allowed us to implement an extreme sub-wavelength EM antenna, radiating an EM wave of wavelength $\lambda = 2$ m, whose emitting area is $\sim 10^{-8}$ m² ( $A/\lambda^2 = 2.5 \times 10^{-9}$), and whose measured radiation efficiency exceeded the $A/\lambda^2$ limit by over $10^5$. The antenna consisted of magnetostrictive nanomagnets deposited on a piezoelectric substrate. A SAW launched in the substrate with an alternating electrical voltage periodically strained the nanomagnets and rotated their magnetizations owing to the Villari effect. The oscillating magnetizations emitted EM waves at the frequency of the SAW. These extreme sub-wavelength antennas, that radiate with efficiencies a few orders of magnitude larger than the $A/\lambda^2$ limit, allow drastic miniaturization of communication systems.

---

[1] This is the pre-review version that was submitted to Advanced Materials Technologies (Wiley). The post-review version that contains small amendments made pursuant to reviewer comments is not posted.

## I. Introduction

Many modern communication systems (cell phones, biologically implanted devices, radio frequency identification systems, wearable electronics) will benefit immensely if antenna dimensions can be reduced to small fractions of the emitting wavelength. Miniaturizing antennas to extreme sub-wavelength dimensions, however, presents a significant challenge because the radiation efficiency is normally limited to $A/\lambda^2$ $(A<\lambda^2)$ where $A$ is the emitting area and $\lambda$ is the wavelength of the emitted radiation [1-5]. One approach to overcoming this limitation is to excite an electromagnetic antenna at acoustic resonance instead of electromagnetic resonance. Since the acoustic wave velocity in many piezoelectric solids is roughly five orders of magnitude smaller than the speed of light in vacuum, the acoustic wavelength is five orders of magnitude smaller than the electromagnetic wavelength at the same frequency. Consequently $A/\lambda_{ac}^2 \sim 10^{10} \times A/\lambda_{EM}^2$, where $A$ is the radiating area of the antenna, $\lambda_{ac}$ is the acoustic wavelength and $\lambda_{EM}$ is the electromagnetic wavelength. This strategy can increase the $A/\lambda^2$ limit on the radiation efficiency of an extreme sub-wavelength electromagnetic antenna by several orders of magnitude, and therefore it has been used to fashion extreme sub-wavelength antennas in the past [6-10]. In particular, ref. [8] demonstrated an electromagnetic antenna, based on this principle, with $A/\lambda_{EM}^2 \approx 10^{-5}$ and an efficiency of ~0.4%. The efficiency exceeded the $A/\lambda_{EM}^2$ limit by a factor of ~400. This work [ref. 8] used a ferromagnetic bulk acoustic wave resonator (FBAR) which was excited acoustically and radiated electromagnetic waves at the resonator's resonance frequency only. In contrast, our antenna is implemented with *many magnetostrictive nanomagnets* whose magnetizations oscillate when subjected to a surface acoustic wave (SAW). The SAW periodically strains the nanomagnets, making their magnetic moments oscillate in time and emit electromagnetic waves. Thus, our antenna works on a different principle and employs a different structure. There are two advantages that our antenna has over the antenna of ref. [8]: First, we can exceed the $A/\lambda_{EM}^2$ limit by a much larger factor (>$10^5$). Second, our antenna is not limited to a particular frequency (e.g. the resonance frequency of an FBAR). As long as the frequency is lower than the inverse spin rotation time of the nanomagnets, it can radiate at any *arbitrary* frequency (which would be the frequency of the SAW launched to activate the nanomagnets). Thus, our antenna can radiate over a wide range of frequencies. All of these antennas are, of course, distinct from conventional antennas where oscillating charges (not oscillating magnetic moments) excite electromagnetic waves.

In this work, we report the experimental demonstration of an extreme sub-wavelength EM antenna, based on the above principle, whose radiation efficiency exceeds the $A/\lambda_{EM}^2$ limit by a factor exceeding $10^5$. The antenna's emitting area is more than 8 orders of magnitude smaller than the square of the wavelength, resulting in drastic miniaturization. It consists of an array of elliptical magnetostrictive (Co) nanomagnets of major axis dimension ~360 nm, minor axis dimension ~330 nm and thickness ~6 nm fabricated on a piezoelectric $128^0$ Y-cut $LiNbO_3$ substrate. A surface acoustic wave (SAW) is launched in the substrate with electrodes and the SAW periodically strains the nanomagnets, causing their magnetizations to rotate owing to the inverse magnetostriction (Villari) effect. The rotating magnetizations emit EM waves (at the frequency of the SAW), which are detected in the far field by a dipole antenna coupled to a spectrum analyzer. The SAW (excitation) frequency in our experiment was 144 MHz and we were able to detect EM emissions at the same frequency that was 8 dBm above ambient emissions, at a distance > 2 m from the antenna. A control sample (that contained no nanomagnets, but was otherwise identical to the actual sample) was used for background subtraction. We were thus able to measure the EM emission from the nanomagnets at the exclusion of all other emitters (e.g. surface currents in the electrodes that are used to launch the SAW and any other spurious source radiating at or near 144 MHz). Based on the measured detected power and the input power to the SAW, we were able to infer that the electromagnetic radiation efficiency (ratio of power radiated by the nanomagnets to the input power to the SAW) is between 0.01% - 0.1%, which exceeds the $A/\lambda_{EM}^2$ limit by more than *five orders of magnitude*.

## II. Results

Figure 1 shows the scanning electron micrographs of the nanomagnet arrays that we fabricated to realize the extreme sub-wavelength antenna. We made two sets of samples: Sample A and Sample B. The former contained 55,000 nanomagnets and the latter 275,000 nanomagnets. Figures 1(a) and 1(b) show low magnification images of several rectangular arrays in the two samples (each white speck is an array) and Fig. 1(c) shows a zoomed image of one such array (where the magnification is not large enough to resolve individual nanomagnets). Figure 2 shows the nanomagnets at higher magnification that allows one to resolve individual nanomagnets and their dimensions.

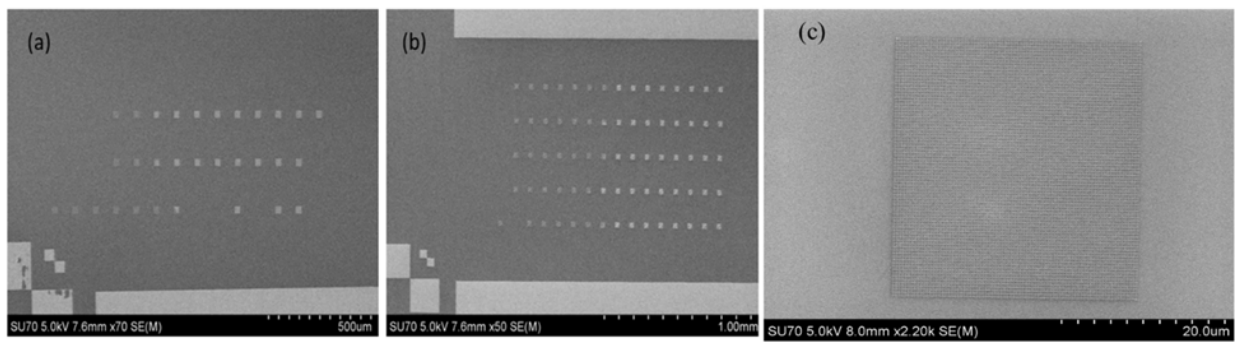

**Fig. 1: Scanning electron micrographs of: (a) nanomagnet arrays in Sample A, (b) nanomagnet arrays in Sample B, and (c) magnified image of one array in Sample B. The magnification is not enough to resolve individual nanomagnets.**

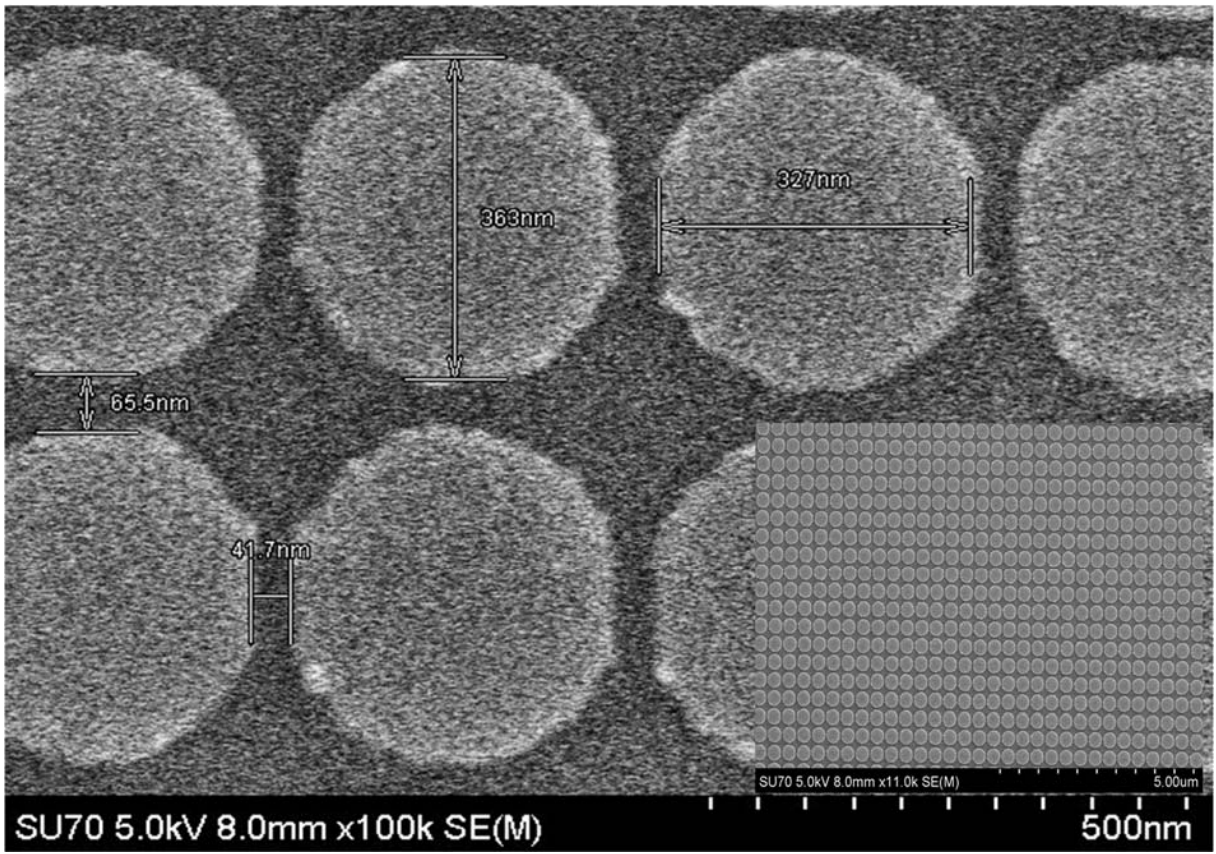

Fig. 2: High magnification scanning electron micrographs of an array in sample B showing that the nanomagnets are slightly elliptical with major axis dimension of ~360 nm, minor axis dimension of 330 nm, vertical edge-to-edge separation of ~65 nm and horizontal edge-to-edge separation of ~42 nm. The inset is a lower magnification image to show the uniformity and density of the array. The nanomagnet thickness is ~6 nm.

We characterized the magnetic behavior of the nanomagnets with static magneto-optical Kerr effect (S-MOKE) at room temperature. Figure 3 shows the Kerr rotation versus magnetic field characteristics (hysteresis loops) under two situations: when the magnetic field is directed along the horizontal axis of the arrays and when directed along the vertical axis. The hysteresis loops confirm that the fabricated nanostructures are ferromagnetic at room temperature. The coercivity is 100 Oe higher when the magnetic field is directed along the horizontal axis which is parallel to the minor axes of the elliptical nanomagnets because this is the hard axis while the major axis is the easy axis.

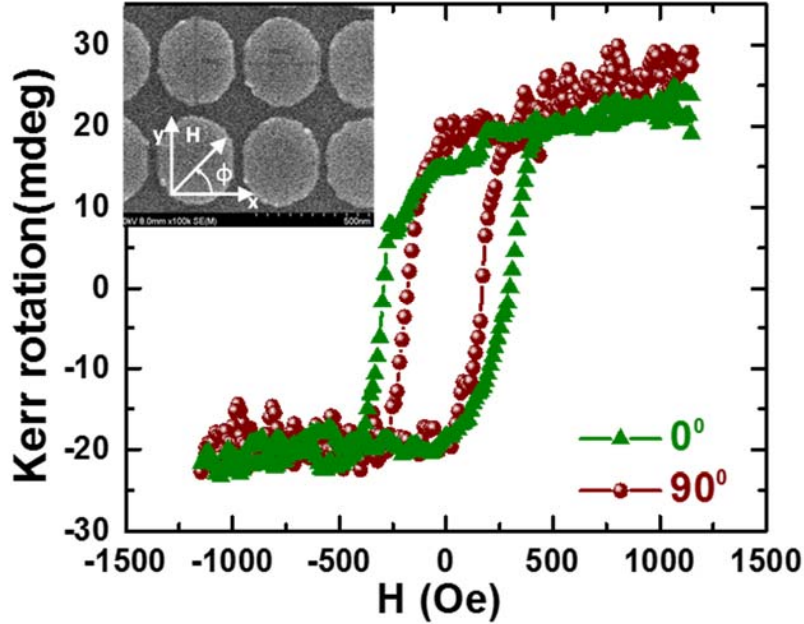

**Fig. 3: Kerr rotation versus magnetic field of the array in Sample B. The Kerr rotation angle is proportional to the magnetic moment. There is ~100 Oe difference between the coercivities when the magnetic field orientation is rotated by $90^0$.**

In order to demonstrate the antenna function and also measure the antenna characteristics, we launched SAW signal in the substrate at two different frequencies ($f_{SAW}$ = 144 MHz and 900 MHz) and measured any detectable EM emission (above the noise floor determined by ambient emissions) at a distance > 2m from the samples. The detector was a dipole antenna calibrated to specific frequencies and these two frequencies belonged to that set, which is why they were chosen. We were able to detect EM emissions at 144 MHz, but not at 900 MHz which was too high a frequency for the magnetization of the nanomagnets to rotate. At 144 MHz, the EM wavelength is 2 m. Since the separation between the detector and the antenna was greater than the EM wavelength, we were measuring the far-field emission.

The SAW velocity in the LiNbO$_3$ substrate is about 4100 m/sec [11], and hence the SAW wavelength is 28.4 μm at 144 MHz, while the EM wavelength is 2 m at that frequency. The ratio of the SAW to EM wavelength is thus $1.42 \times 10^{-5}$.

We carried out the measurements for both samples A and B containing nanomagnets, as well as control samples that were otherwise identical to the real samples but had no nanomagnets. In Fig. 4, we show the detection results at $f_{SAW}$ = 900 MHz for Sample A (screenshots of the spectrum analyzer are shown in the Supplementary Material). There is ~1 dB difference between the 900 MHz emissions from the real sample (with nanomagnets) and the control sample (without nanomagnets), indicating that the nanomagnets are radiating very weakly at 900 MHz, if at all. Most likely, the nanomagnets do not radiate sufficiently because they are not able to rotate their magnetizations through sufficiently large angles at this high rate (900 MHz). Surface currents induced in the electrodes used to launch the SAW in both samples also radiate electromagnetic waves at 900 MHz and the detected emissions are primarily due to that. Similar (negative) result was obtained with Sample B.

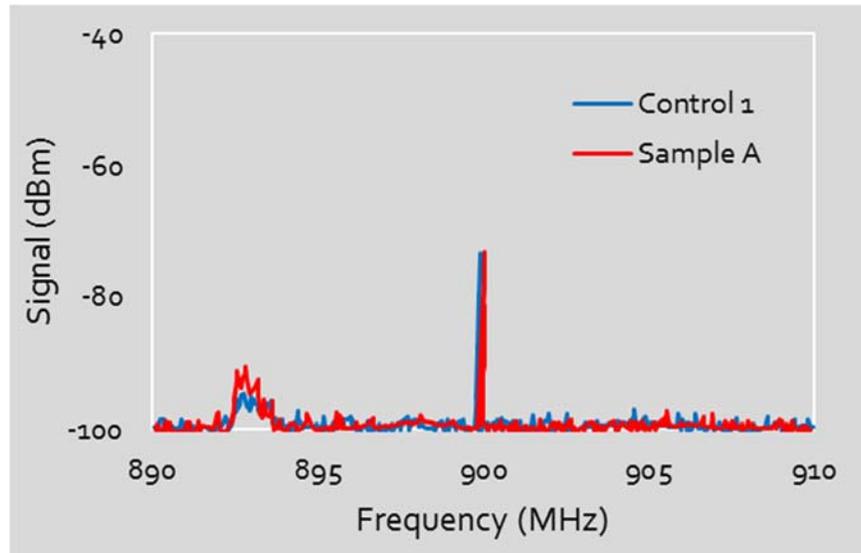

Fig. 4: The power spectra of electromagnetic emissions detected by a dipole antenna coupled to a spectrum analyzer when the SAW frequency is 900 MHz. The antenna is placed > 2 m from the sample. The spectra are shown for Sample A and the control sample. There is only ~1 dB difference between the detected electromagnetic powers from the two samples showing that the nanomagnets are not radiating sufficiently at this frequency. Screenshots of the spectrum analyzer can be found in the Supporting Information. The input power to launch the SAW was 5 dbm (3.16 mW).

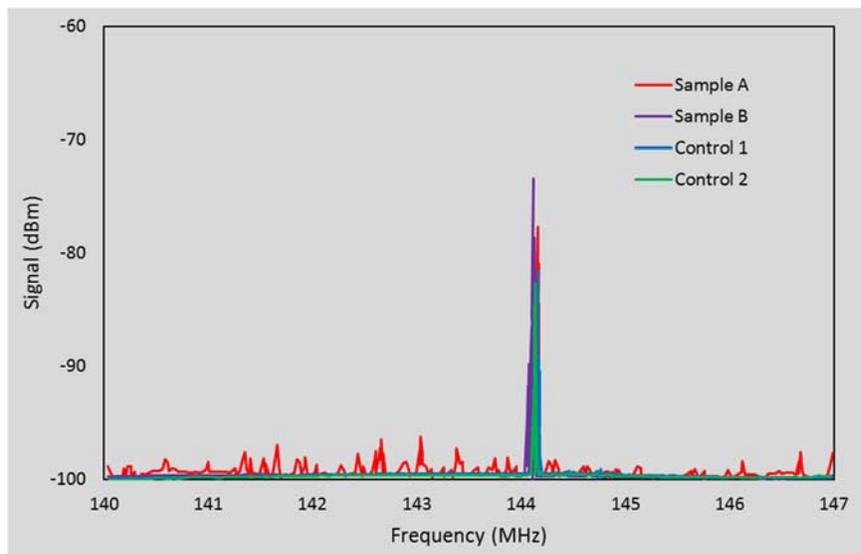

Fig. 5: The power spectra of emissions detected when the SAW frequency is 144 MHz. Control 1 is the control sample corresponding to Sample A, and Control 2 is the one corresponding to Sample B. The input power was 5 dbm (3.16 mW). The detected emission from Sample B is 8 dB above that of Control 2 and is -73 dbm (50 pW).

Our past simulations have shown that strain induced large angle magnetization rotation in single domain elliptical nanomagnets (of lateral dimension ~100 nm) typically takes place in about 1 ns [12-14]. The nanomagnets used here are larger (> 300 nm lateral dimension) and multi-domain. Hence, it is possible that they rotate slower and therefore, the period of the 900 MHz signal (1.1 ns) does not allow them enough time to rotate through a large angle and radiate electromagnetic waves. Subsequently, we reduced the excitation frequency to 144 MHz and the detection results are shown in Fig. 5. We now clearly see a measurable difference between the real samples and the control samplse. The detected radiation power from Sample B is 8 dB higher than that from the control sample, while that from Sample A is 3 dB higher than that from the control sample. These differences indicate that the nanomagnets are able to rotate their magnetizations at this lower frequency and emit EM waves.

We also carried out time-resolved magneto-optical Kerr effect (TR-MOKE) measurements on the nanomagnets at room temperature at various amplitudes of SAW excitation to verify that the launched SAW indeed has an effect on the magnetization rotation. The oscillations in time-resolved Kerr rotations are measured with a micro-focused optical pump-probe set up as shown in Fig. 6(a). Details of the set-up (e.g. beam spot size, pulse width, repetition rate, etc.) can be found elsewhere [15, 16] and hence not repeated here. The measurements were done in the absence of any bias magnetic field. The ultrashort laser pulses used in the TR-MOKE measurements set up very high frequency (~ 4 GHz) oscillations of the nanomagnets' magnetizations [15] and surprisingly, we found that the amplitudes of the Kerr oscillations resulting from these high frequency oscillations are significantly increased by the launched SAW with $f_{SAW}$ = 144 MHz. The amplitudes are markedly different in the absence of SAW versus in the presence of SAW. The amplitudes also show a rather weak dependence on the launched SAW power ($P$) for $P$ > -15 dBm.

These results are shown in Figs. 6(b) – 6(c). Note that the SAW frequency ($f_{SAW}$ = 144 MHz) is more than an order of magnitude lower than the Kerr oscillation frequencies which are in the neighborhood of 4 GHz. The Kerr oscillations are not caused by the launched SAW. Instead, they are caused by the ultrashort laser pulses in the TR-MOKE set-up. The excitation by the femtosecond laser causes an ultrafast demagnetization of the nanomagnets followed by two-step relaxation (not shown) which also launches an ultrafast internal field to trigger magnetization precession of the nanomagnets. The absence of any bias magnetic field ensures that the magnetization precesses around an effective magnetic field due to the dipolar coupling fields between the nanomagnets, which leads to a dominant natural resonance frequency at around 4 GHz. Clearly the launched SAW strongly affects the amplitude of this resonant oscillation of magnetization despite being highly off-resonant and having a very weak SAW power. In Fig. 6(d), we show the fast Fourier transforms (power spectral densities) of the Kerr oscillations for various SAW power. The ensuing power spectral densities are also affected by the SAW due to the variation in the magnetization oscillation amplitudes.

In the TR-MOKE experiments, we cannot detect any Kerr oscillation having a frequency component at the launched SAW frequency of 144 MHz because the time delay between the pump and probe laser (Δt) is only up to 3 ns and hence the lowest frequency component that can be resolved is about 1/3 ns, i.e. 333 MHz. Therefore, we further launched a SAW of frequency $f_{SAW}$ = 350 MHz. The time-resolved Kerr rotations and their fast Fourier transforms are shown in Fig. 7(a) and 7(b). In Fig. 7(b), we are still not able to detect any clear and consistent peak at 350 MHz. This indicates that either the SAW, by itself, cannot induce sufficient magnetization rotation at this high frequency of 350 MHz, or any signature of that rotation is being drowned by the much stronger Kerr oscillations caused not by the SAW, but by the ultrashort laser pulses. However, the clear observation that the SAW power affects Kerr oscillations significantly indicates unambiguously that the SAW affects magnetization oscillations and thus could be the source of the electromagnetic emission which we observe.

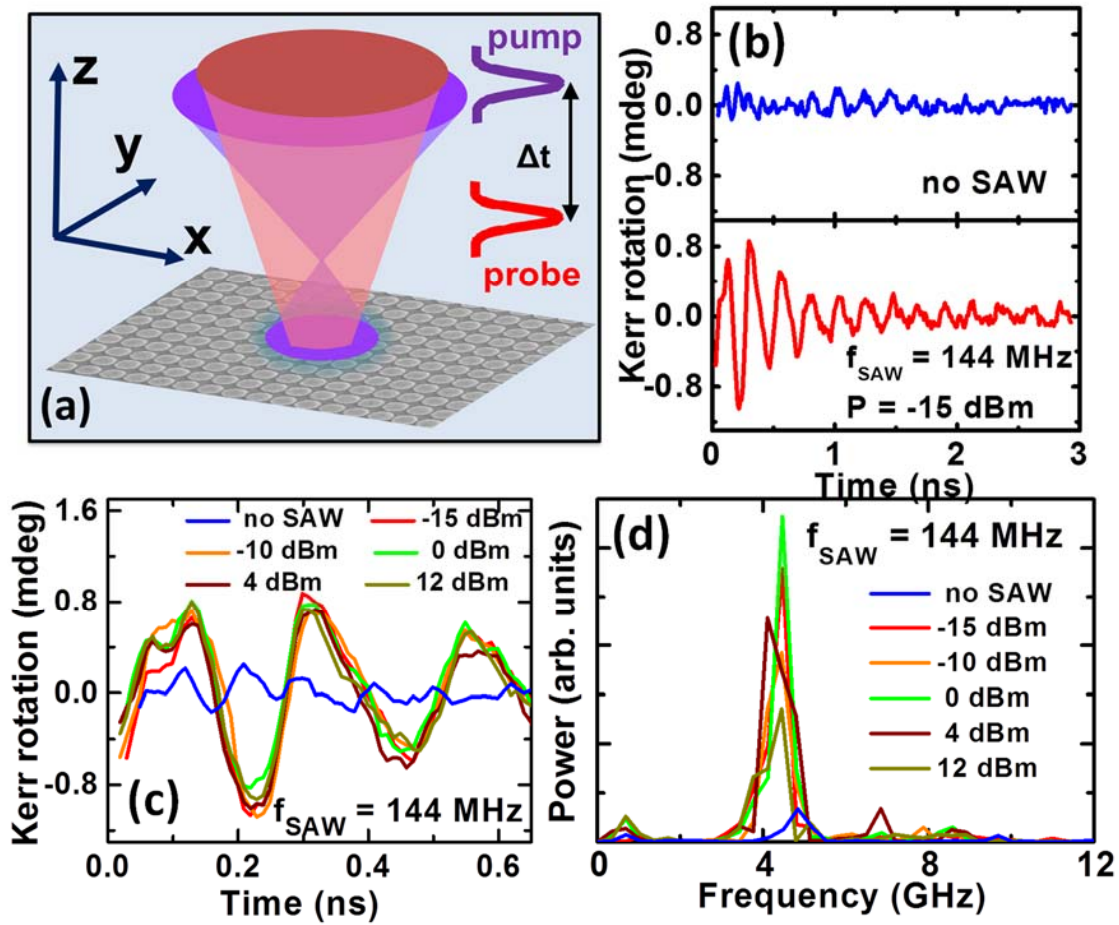

Fig. 6: (a) Schematic of the TR-MOKE microscope showing the pump and probe beams and the geometry of the measurement. The probe beam spot covers about 4 nanomagnets. (b) The oscillations in time-resolved Kerr rotation in the absence of any externally launched SAW (top panel) and in the presence of an externally launched SAW at 144 MHz frequency with a power of -15 dBm (31.6 µW). The ultrashort laser pulses however independently induce high frequency SAW waves in the substrate, which cause high frequency oscillation of the magnetizations, leading to some Kerr oscillations. The oscillations are much more pronounced and have much larger amplitudes (before damping sets in) in the presence of the SAW. (c) The oscillations in time-resolved Kerr rotations at various SAW power levels. The SAW frequency is 144 MHz, which is the frequency used in the antenna experiment. (d) (a) Power spectral densities of the Kerr oscillations at launched SAW frequency of 144 MHz for various SAW power.

### III. Calculation of antenna radiation efficiency

In Fig. 8, we show the reflection coefficient $S_{11}$ (measured at the electrodes that launch the SAW) at the input power of 5 dbm (3.16 mW) as a function of frequency. The measurements are carried out with a network analyzer for Sample A, as well as the control samples. At the frequency of 144 MHz, ~85% of the input power is reflected back to the source because of impedance mismatch and hence only about 15% of the input power is available to be coupled into the SAW. Therefore, the maximum actual power fed to the antenna via the SAW is $3.16 \times 0.15 = 0.474$ mW. This is the input power to Sample A, and the input power to Sample B is about the same.

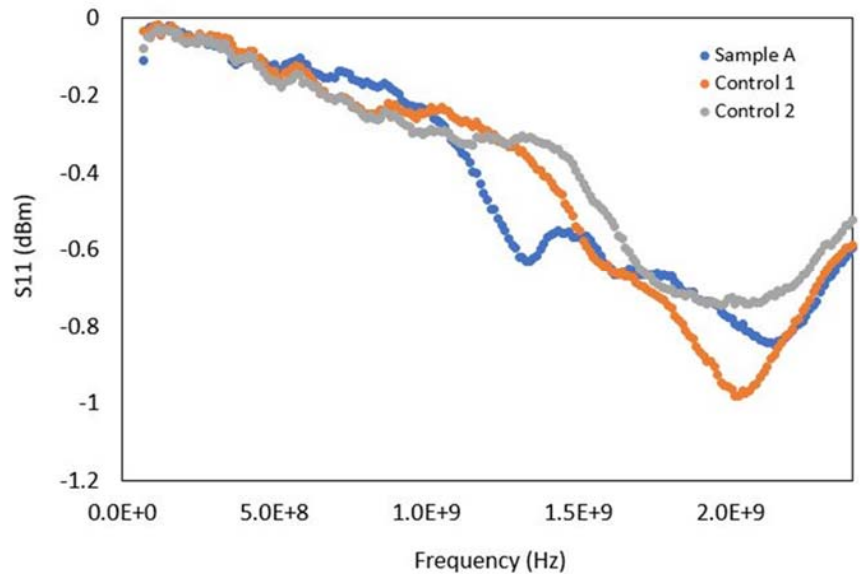

**Fig. 8: Reflection coefficient $S_{11}$ as a function of input signal frequency.**

The EM power from Sample B that is detected by the receiving dipole antenna is -73 dBm (see Fig. 5) which is about 50 pW. The power radiated by the control sample that is detected by the antenna is -81 dBm, which is 8 pW. Hence the power detected from the nanomagnets is 42 pW. The actual radiated power from Sample B is radiated over $4\pi$ solid angle and the fraction that is incident on the receiving dipole antenna (and hence detected) is $\frac{lw}{4\pi r^2}$ where $l$ is the length of the receiving antenna, $w$ is its width and $r$ is the separation between the source and the detector. In our case $l = 1$ m, $w = 0.5$ cm and $r = 2$ m. Hence the ratio $\frac{lw}{4\pi r^2}$ is $10^{-4}$ and consequently, the power actually radiated by Sample B (over $4\pi$ solid angle) is $42 \times 10^4$ pW $= 0.42$ µW. Consequently, the radiation efficiency, which is the ratio of the radiated power to the input power, is 0.42 µW/0.474 mW = 0.088% in the case of Sample B. In the case of Sample A, the detected power was -78 dBm, which is about 15 pW. Therefore, the power from the nanomagnets in sample A that was detected by the antenna was 15 pW – 8 pW = 7 pW. The radiated power was hence $7 \times 10^4$ pW = 0.07 µW. In this case, the efficiency is 0.07 µW/0.474 mW = 0.014%. Since Sample A had 55,000 nanomagnets and Sample B had 275,000 nanomagnets, we expect the radiation from Sample A to be weaker than that from Sample B, and we observed that to be the case.

Let us now calculate the $A/\lambda_{EM}^2$ limit for both samples. The area of a nanomagnet is $\frac{\pi}{4}(a \times b)$ where $a$ is the major axis dimension (360 nm) and $b$ is the minor axis dimension (330 nm). Since there are 55,000 nanomagnets in Sample A, the radiating area is $A = \frac{\pi}{4}(360 \times 330) \times 10^{-18} \times 55,000 = 5 \times 10^{-9}$ m². Hence, in the case of Sample A, $\left.\frac{A}{\lambda_{EM}^2}\right|_{Sample\ A} = 1.25 \times 10^{-9}$, which means that our measured efficiency of 0.014% was able to beat the $A/\lambda_{EM}^2$ limit by 112,000

times. In the case of Sample B, the radiating area is $A = \frac{\pi}{4}(360 \times 330) \times 10^{-18} \times 275,000 = 2.5 \times 10^{-8}$ m². Hence for Sample B, $\left.\frac{A}{\lambda_{EM}^2}\right|_{\text{Sample B}} = 6.25 \times 10^{-9}$, which means that our measured efficiency of 0.088% was able to beat the $A/\lambda_{EM}^2$ limit by 140,800 times in Sample B.

## IV. Conclusion

In this work, we have demonstrated extreme sub-wavelength electromagnetic antennas whose radiation efficiencies greatly exceed the theoretical limit of $A/\lambda_{EM}^2$ $(A<\lambda_{EM})$ [where $A$ is the emitting area and $\lambda_{EM}$ is the wavelength of the emitted electromagnetic wave] by a factor exceeding $10^5$. This allows us to miniaturize electromagnetic antennas. In our case, the emitting areas of the antennas are about $2 \times 10^8$ times smaller than the square of the emission wavelength. This drastic miniaturization was made possible by exciting the antennas at acoustic resonance instead of electromagnetic resonance. The surface acoustic waves were also found to amplify the magnetization response of these nanomagnets resonating in GHz frequencies, which may have different applications of its own. These extreme sub-wavelength antennas will allow dramatic downscaling of communication systems and could open up new high frequency applications.

## V. Experimental Details

The nanomagnets were fabricated on a 128° Y-cut LiNbO$_3$ substrate. The substrate was spin-coated with bilayer polymethyl methacrylate (PMMA) e-beam resists of different molecular weights to obtain good undercut: PMMA 495 diluted 4% by volume in Anisole, followed by PMMA 950 also diluted 4% by volume in Anisole. The spin coating was carried out at a spinning rate of 2500 rpm. The resists were subsequently baked at 110° C for 5 min. Next, electron-beam lithography was performed using a Hitachi SU-70 scanning electron microscope (at an accelerating voltage of 30 kV and 60 pA beam current) with a Nabity NPGS lithography attachment. Finally, the resists were developed in methyl isobutyl ketone and isopropyl alcohol or MIBK−IPA (1:3) for 270 s followed by a cold IPA rinse.

For nanomagnet delineation, a 5-nm-thick Ti adhesion layer was first deposited on the patterned substrate using e-beam evaporation at a base pressure of ∼2 × 10$^{-7}$ Torr, followed by the deposition of Co. The lift-off was carried out using Remover PG solution.

## ASSOCIATED CONTENT

### Supporting Information

Supporting information is available free of charge from the ACS Publication Website. It contains two figures:

S1: Screenshot of spectrum analyzer when SAW frequency was 900 MHz.

S2: Screenshot of spectrum analyzer when SAW frequency was 144 MHz.

and an estimation of the minimum coupling efficiency of the magnetic moments to electromagnetic radiation.

## AUTHOR INFORMATION

### Corresponding author:

*Email: (S. B.) sbandy@vcu.edu

### Author Contributions

J. L. D fabricated all the samples and imaged them with scanning electron microscopy; A. D., K. D. and P. P. carried out the time-resolved magneto-optical Kerr effect measurements and analyses and A. A. performed the static MOKE measurements under the supervision of A. B. J. L. D. and J. L. carried out the electromagnetic measurements under the supervision of E. T; S. B. conceived of the experiments, supervised all measurements, analyzed the data, and wrote the manuscript, which was critiqued by all authors.

## ACKNOWLEDGEMENT

The authors are deeply indebted to Mr. Jonathan Lundquist for his help with obtaining the electromagnetic emission data and to Prof. Erdem Topsakal for his helpful suggestions with regard to the electromagnetic experiments. The work at Virginia was supported by a Virginia Commonwealth University Commercialization grant. Both institutions received support from an Indo-US Science and Technology Fund Center grant titled "Center for Nanomagnetics for Energy-Efficient Computing, Communications and Data Storage" (IUSSTF/JC-030/2018).

**Conflict of Interest**

The authors declare no conflict of interest.

**References:**

1. B. A. Kramer, C.-C. Chen, M. Lee and J. L. Volakis, Fundamental limits and design guidelines for miniaturizing ultra-wideband antennas, *Antennas Propag. Mag. IEEE*, **51**, 57–69 (2009).

2. H. Mosallaei and K. Sarabandi, Antenna miniaturization and bandwidth enhancement using a reactive impedance substrate, *IEEE Trans. Antennas Propag.* **52**, 2403–2414 (2004).

3. A. K. Skrivervik, J. F. Zürcher, O. Staub and J. R. Mosig, PCS antenna design: The challenge of miniaturization, *IEEE Antennas Propag. Mag.*, **43**, 12–27 (2001).

4. J. P. Gianvittorio and Y. Rahmat-Samii, Fractal antennas: a novel antenna miniaturization technique, and applications, *IEEE Antennas Propag. Mag.*, **44**, 20–36 (2002).

5. P. M. T. Ikonen, K. N. Rozanov, A. V. Osipov, P. Alitalo and S. A. Tretyakov, Magnetodielectric substrates in antenna miniaturization: potential and limitations, *IEEE Trans. Antennas Propag.*, **54**, 3391–3399 (2006).

6. Z. Yao, Y. E. Wang, S. Keller and G. P. Carman, Bulk acoustic wave mediated multiferroic antennas: architecture and performance bound, *IEEE Trans. Antennas Propag.*, **63,** 3335–44 (2015).

7. J. P. Domann and G. P. Carman, Strain powered antennas, *J. Appl. Phys.*, **121,** 044905 (2017).

8. T. Nan, H. Lin, Y. Gao, A. Matyushov, G. Yu, H. Chen, N. Sun, S. Wei, Z. Wang, M. Li, X. Wang, A. Belkessam, R. Guo, B. Chen, J. Zhou, G. Qian, Y. Hui, M. Rinaldi, M. E. McConney, B. M. Howe, Z. Hu, J. G. Jones, G. J. Brown and N. X. Sun, Acoustically actuated ultra-compact NEMS magnetoelectric antennas, *Nat. Commun.* **8***,* 296 (2017).

9. H. Lin, M. Zaeimbashi, N. Sun, X. Liang, H. Chen, C. Dong, A. Matyushov, X. Wang, Y. Guo, and N. X. Sun, NEMS magnetoelectric antennas for biomedical applications, *IMBioc 2018–2018 IEEE/MTT-S Int. Microwave Biomedical Conf.*, (IEEE, Piscataway, New Jersey) 13–15 (2018).

10. H. Lin, M. R. Page, M. McConney, J. Jones, B. Howe and N. X. Sun, Integrated magneto-electric devices: filters, pico-Tesla magnetometers, and ultra-compact acoustic antennas, *MRS Bull.*, **43**, 841–847 (2018).

11. A. Holm, Q. Stürzer, Y, Xu and R. Weigel, Investigation of surface acoustic waves on LiNbO$_3$, quartz, and LiTaO$_3$ by laser probing, *Microelectronic Engineering*, **31**, 123 (1996).

12. M. Salehi-Fashami, K. Roy, J. Atulasimha and S. Bandyopadhyay, Magnetization dynamics, Bennett clocking and associated energy dissipation in multiferroic logic, *Nanotechnology*, **22**, 155201 (2011).

13. K. Roy, S. Bandyopadhyay and J. Atulasimha, Energy dissipation and switching delay in stress-induced switching of multiferroic nanomagnets in the presence of thermal fluctuations, *J. Appl. Phys.*, **112**, 023914 (2012).

14. A. K. Biswas, S. Bandyopadhyay and J. Atulasimha, Complete magnetization reversal in a magnetostrictive nanomagnet with voltage generated stress: A potential non-volatile random access memory, *Appl. Phys. Lett.*, **105**, 072408 (2014).

15. S. Mondal, M. A. Abeed, K. Dutta, A. De, S. Sahoo, A. Barman and S. Bandyopadhyay, Hybrid magneto-dynamical modes in a single magnetostrictive nanomagnet on a piezoelectric substrate arising from magnetoelastic modulation of precessional dynamics, *ACS Appl. Mater. Interfaces*, **10**, 43970-43977 (2018).

16. A. Barman and J. Sinha in *Spin Dynamics and Damping in Ferromagnetic Thin Films and Nanostructures*, Springer International Publishing AG, (2018).

# Supplementary Material:
# Extreme sub-wavelength magneto-elastic electromagnetic antenna implemented with multiferroic nanomagnets

J. L. Drobitch[1], A. De[2], K. Dutta[2], P. Pal[2], A. Adhikari[2], A. Barman[2] and S. Bandyopadhyay[1]

[1]Dept. of Electrical and Computer Engineering, Virginia Commonwealth University, Richmond, VA 23284, USA

[2]Dept. of Condensed Matter Physics and Material Science, S. N. Bose National Center for Basic Sciences, JD Block, Sector III, Salt Lake, Kolkata 700 106, India

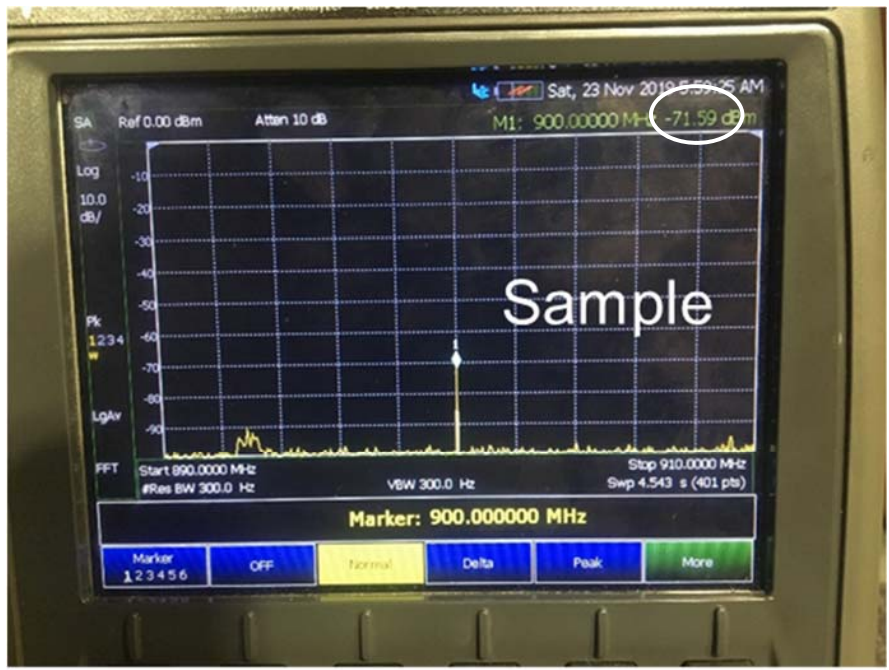

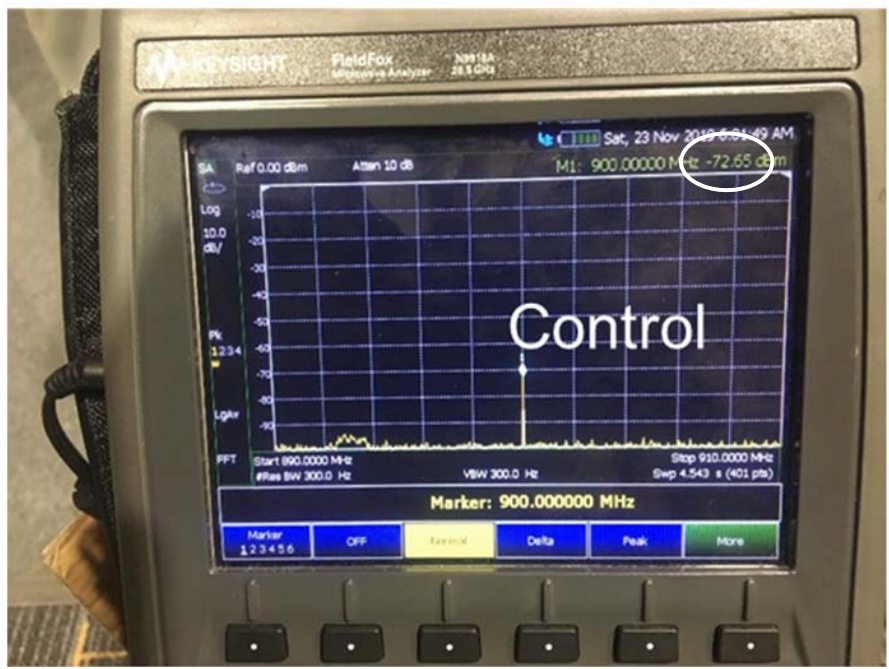

Fig. S1: Screenshots of the power spectra of emissions detected from Sample A and the control sample when the SAW frequency was 900 MHz. The input power was 5 dbm. The detecting antenna was placed at a distance > 2 m from the sample. There is slightly less than 1 dB difference between the power detected from Sample A and the control sample, showing that the nanomagnets in Sample A are not radiating sufficient power. Similar small difference was also observed in the case of Sample B. The detected emissions in this case are primarily due to surface currents induced in the electrodes used to launch the SAW. These surface currents also radiate at 900 MHz.

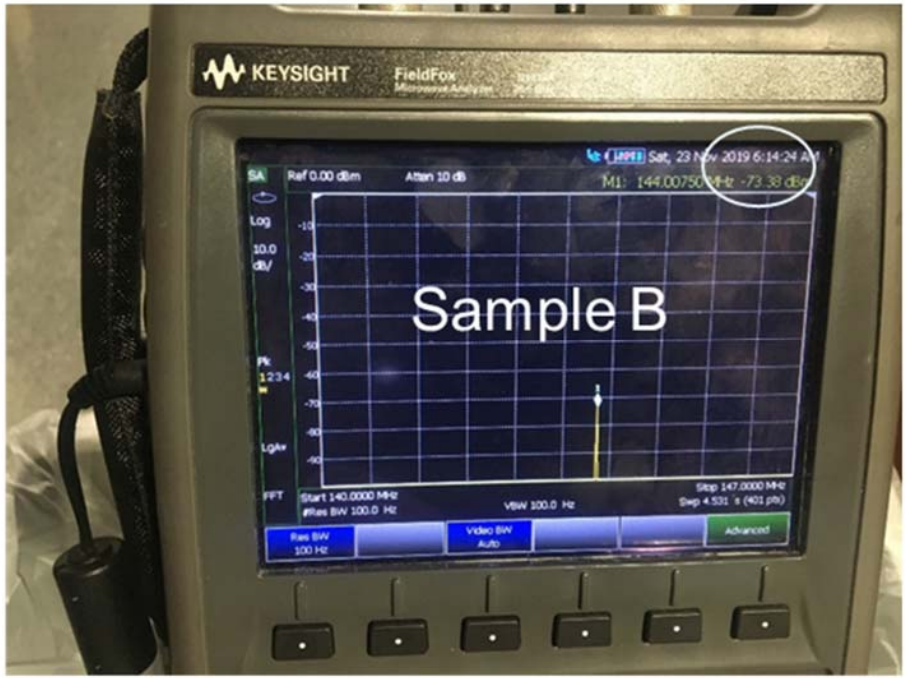
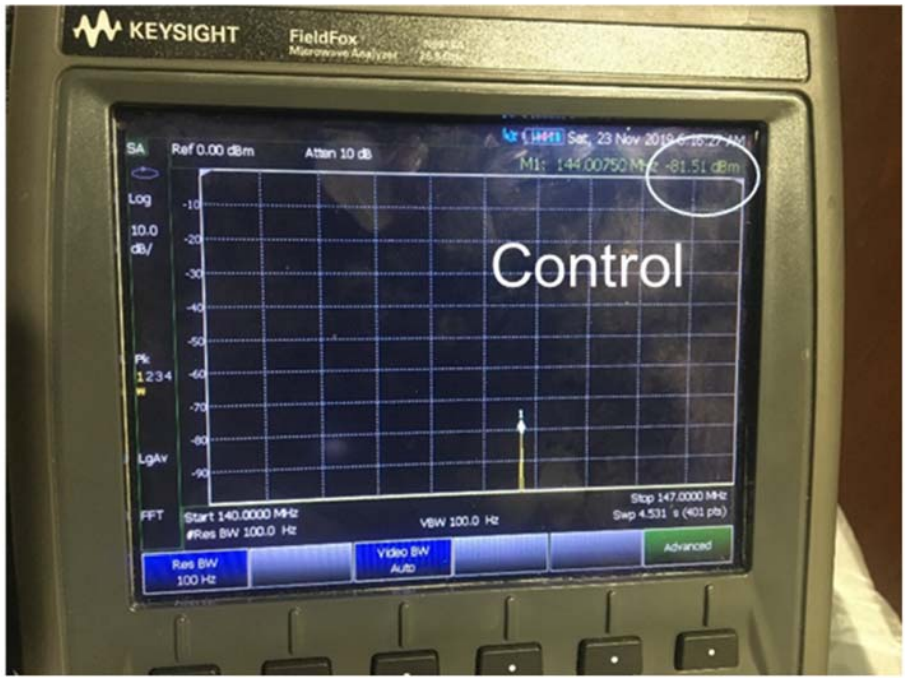

Fig. S2: Screenshots of the power spectra of emissions detected from Sample B and the corresponding control sample when the SAW frequency was 144 MHz. The input power was 5 dbm. The detecting antenna was placed at a distance > 2 m from the sample. There is more than 8 dB difference between the power detected from Sample B and the control sample at this lower frequency, showing that the nanomagnets in Sample B are radiating much more strongly than at 900 MHz.